\documentclass[prd,aps,twocolumn,%
showpacs,preprintnumbers,amsmath,amssymb]{revtex4-2}
\usepackage[dvipdfmx]{graphicx}
\usepackage[breaklinks, colorlinks=true]{hyperref}
\usepackage{bm}

\usepackage{comment}

\newcommand{\Tr}{\mathop{\mathrm{Tr}}}
\newcommand{\Nf}{N_{\mathrm{f}}}
\newcommand{\Nc}{N_{\mathrm{c}}}
\newcommand{\QCtwo}{\mathrm{QC_2D}}
\newcommand{\muiso}{\mu_{\mathrm{I}}}
\newcommand{\muq}{\mu_{\mathrm{q}}}

\begin{document}

\title{Speed of sound and trace anomaly in a unified treatment of the two-color diquark superfluid, the pion-condensed high-isospin matter, and the 2SC quark matter}
\date{\today}

\author{Kenji Fukushima}
\email{fuku@nt.phys.s.u-tokyo.ac.jp}
\affiliation{Department of Physics, The University of Tokyo,
             7-3-1 Hongo, Bunkyo-ku, Tokyo 113-0033, Japan}

\author{Shuhei Minato}
\email{minato@nt.phys.s.u-tokyo.ac.jp}
\affiliation{Department of Physics, The University of Tokyo,
             7-3-1 Hongo, Bunkyo-ku, Tokyo 113-0033, Japan}

\begin{abstract}
    In a unified perturbative treatment from the high-density side, we compute the speed of sound and the trace anomaly as functions of the chemical potential $\mu$ for the two-color diquark superfluid, the pion-condensed high-isospin matter, and the two-flavor color superconducting (2SC) quark matter.
    We find that the corrections induced by the gap energy $\Delta$ involve nontrivial interplay between the dimensionless magnitude $|\Delta/\mu|$ and the derivative $|\partial\Delta/\partial\mu|$.  Even though the gap equation has a common structure for these phases of our interest, different numerical constants cause drastic changes in the speed of sound corrections.  As long as $|\Delta/\mu|$ is dominant over the derivative, the gap effects increase the speed of sound, which is consistent with the expected behavior of exhibiting a peak at intermediate density.
    We then discuss the trace anomaly which is pushed down generally by the gap effects and turns out to be negative widely in the high density regime.
    For demonstration of non-perturbative enhancement, we take account of the instanton-induced interaction.
    We confirm a further increase in the speed of sound for the two-color diquark superfluid and the pion-condensed high-isospin matter, while the speed of sound for the 2SC quark matter deviates far below the conformal limit due to the derivative contribution.
\end{abstract}
\maketitle

\section{Introduction}

Quantum chromodynamics (QCD) at finite density is a long standing problem which suffers the notorious sign problem in the first-principles lattice simulation~\cite{Troyer:2004ge, Nagata:2021ugx}.
In principle, QCD should predict a relation between the pressure $p$ and the internal energy density $\epsilon$ of strongly interacting matter.  This relation called the equation of state is indispensable for the understanding of the structures of neutron stars (NSs).  This in turn means that the observational data from the NSs can constrain the QCD equation of state.  In this context, the most well-known example is the constraint from the existence of the two-solar-mass NS that has excluded scenarios with soft equations of state~\cite{Demorest:2010bx}.

The stiff equation of state implies a larger $p$ for $\epsilon$ or a larger slope $dp/d\epsilon$, i.e., the speed of sound $c_s$.  As explained later, the perturbative QCD (pQCD) yields $c_s^2<1/3$ at high density and the conformal limit, $c_s^2=1/3$, is achieved asymptotically.  Because the speed of sound in nuclear matter is far below the conformal limit, it may be a plausible conjecture that $c_s^2 \le 1/3$ should hold for any density.

The recent analysis of the NS observation data, however, supports a stiff equation of state in the intermediate density where the speed of sound has a peak exceeding the conformal value.~\cite{Masuda:2012ed, Fujimoto:2019hxv, *Fujimoto:2021zas, Altiparmak:2022bke, Brandes:2023hma}  The microscopic origin of the speed of sound peak is still under debate and various models have been proposed so far~\cite{Baym:2017whm,McLerran:2018hbz}.  One scenario is that the speed of sound peak signals for a transition from nuclear to quark matter.  Usually, a phase transition suppresses the speed of sound, and the transitional change from nuclear to quark matter should take place in an unconventional way.  To enhance $dp/d\epsilon$, the pressure should be increased for a relatively small change in the energy density.  This may be possible if nucleons and quarks are interchanged inside the Fermi sea --- the pressure is increased while the baryon number is intact.  This picture is referred to as the momentum-shell model in quarkyonic matter.  The quarkyonic matter model has been refined as a solvable model in which the nucleon-quark duality is manifest \cite{Fujimoto:2023mzy}.

The peak behavior of the speed of sound is characterized by the trace anomaly, which is denoted by $\mathfrak{T}$ in this paper to discriminate from the gap energy $\Delta$.  In the context of the NS data analysis, $\mathfrak{T}$ was introduced in \cite{Fujimoto:2022ohj}.  There, it has been shown that the rapid recovery of conformality triggers a peak in $c_s^2$.  According to later studies, it is likely that $\mathfrak{T}$ may cross the zero at some density and goes negative.  Once $\mathfrak{T}$ becomes negative, the thermal degrees of freedom must decrease with increasing chemical potential, which is counter-intuitive.  Nevertheless, this exotic behavior is expected if the system accommodates nonzero condensates.

Stimulating results have been derived from the first-principles lattice simulations in analogous setups.
The first notable demonstration with a QCD-like theory we introduce is the speed of sound calculation in dense two-color QCD, i.e., dense $\QCtwo$.  Because $SU(2)$ is pseudo-real, dense $\QCtwo$ is free from the sign problem.  Moreover, the diquark is a color-singlet in $\QCtwo$, and the superfluid phase has a finite diquark condensate with spontaneous $U(1)_B$ breaking.  The most attractive two-color diquark is spin-antisymmetric (singlet,) flavor-antisymmetric, color-antisymmetric, and spatially-symmetric. These superfluid condensates are reminiscent of the color superconducting condensates in high density QCD\@.  It is clearly seen in Refs.~\cite{Iida:2022hyy, Iida:2024irv} that $c_s^2$ exceeds $1/3$ around $\muq\simeq 0.6 m_{\mathrm{PS}}$ where $\mu_{q}$ is the quark chemical potential and $m_{\mathrm{PS}}$ is the diquark (pion) mass in the pseudo-scalar channel.

Another remarkable analogue is the high-isospin matter~\cite{Kojo:2024sca}.  When the isospin chemical potential $\muiso$ goes above the pion mass, the pion condensation forms~\cite{Son:2000xc, Kogut:2002zg}.  The chiral effective theory (ChEFT) gives a reliable estimate of the thermodynamic quantities at low energies and the speed of sound approaches the unity. Based on ChEFT, $p\simeq \varepsilon$ is concluded in the high density limit, and accordingly $\mathfrak{T}$ gets negative. Since the ChEFT is valid at low energies, the speed of sound cannot be the unity but should go to the conformal value in the physical high density limit.  A recent lattice QCD simulation has confirmed that $c_s^2$ certainly exceeds the conformal value around $\muiso \simeq 1.5m_\pi$~\cite{Brandt:2022hwy, Abbott:2023coj}.  Interestingly, the trace anomaly has also been measured and $\mathfrak{T} < 0$ is realized there.

The most challenging problem is to disclose the origin of $\mathfrak{T}\lesssim 0$ for the NS matter.  The natural candidate for nonzero condensation is the color superconductivity \cite{Rajagopal:2000wf,Alford:2007xm}.  The color-antisymmetric channel is perturbatively attractive, and the spin-antisymmetric (singlet) and the flavor-antisymmetric (anti-triplet) combination of two quarks constitutes the ``good'' diquarks.  In flavor space there are three distinct diquarks.  When these three are all finite (and degenerated), the diquark condensates characterize the color-flavor-locked (CFL) phase.  As $\muq$ gets lowered, the Fermi surface mismatch between the light flavors and the strangeness breaks the paring with $s$ quarks, and only the $u$-$d$ diquark condensate survives, which is nothing but the two-flavor color superconducting (2SC) phase.

 In this paper, we investigate the speed of sound and the trace anomaly by means of the perturbative calculation.  Without condensates, the na\"{i}ve perturbative estimate seems incompatible with the lattice results.  It is intriguing to check how the condensates in the two-color diquark superfluid, the pion-condensed high-isospin matter, and the 2SC quark matter would affect the speed of sound and the trace anomaly quantitatively~\cite{Leonhardt:2019fua}.
Indeed, for the normal-phase quark matter, the next-to-next-to-leading order (N$^2$LO) calculation of thermodynamic pressure in the chiral limit was completed about a half century ago~\cite{Freedman:1976ub, Beluni:1978}.  The finite quark mass was taken into account~\cite{Kurkela:2009gj}.  Regardless of the quark mass, the N$^2$LO calculation predicts $c_s^2 < 1/3$ at high density.

The pQCD calculation for the systems with above-mentioned condensates has been established.  The condensate or the gap energy is provided as the analytical solution of the one-loop gap equation~\cite{Son:1998uk, Pisarski:1999tv, Brown:1999aq, Wang:2001aq}.  In Refs.~\cite{Alford:2007xm, Kurkela:2009gj}, the gap effects on the pressure has been discussed for the given analytical form.
In the present paper, we will update this program of including the gap effects in thermodynamics in two ways.
First, we find that a factor difference in the gaps would cause drastic changes in the speed of sound, and we try to get rid of approximations introduced to simplify the analytical solution.  Although those approximations are reasonable at sufficiently high density, our numerically solved gaps are factor larger than the analytical ones, which has a significant impact on the speed of sound.
Second, the diquark gap correction to the pressure has been estimated by only the leading order (LO) calculation~\cite{Alford:2007xm, Kurkela:2009gj}, i.e., $\mathcal{O}(\mu^2\Delta^2)$.  The next corrections should be of order of $g^2\mu^2\Delta^2$ as well as $g^2\mu^2\Delta^2 \ln(m_D/\Delta)\sim g\mu^2\Delta^2$.  For the latter correction, the exponent of $\Delta$ cancels $g$ in the overall constant.
Recently, $\mathcal{O}(g^2\mu^2\Delta^2)$ corrections from the sun-set diagram~\cite{Geissel:2024nmx} have been considered, and also a part of $\mathcal{O}(g\mu^2\Delta^2)$ corrections from the fermionic one-loop diagram has been studied~\cite{Fujimoto:2023mvc}. 
In this work, we incorporate the $\mathcal{O}(g\mu^2\Delta^2)$ corrections with not only the fermionic but also the gluonic one-loop resummed diagram (with insertion of fermionic loops).
It should be noted that in the color-superconducting phase this gluonic one-loop diagram gives rise to the $\mathcal{O}(g\mu^2\Delta^2)$ correction, which originates from the Meissner mass~\cite{Evans:1999at}. We will address details about this gluonic contribution in the main text within a suitable approximation and investigate how the speed of sound is affected.

We emphasize that, as we will perceive from the expressions shown later, the underlying physical mechanisms for the two-color diquark superfluid, the pion-condensed high-isospin matter, and the 2SC quark matter are common.  Therefore, it is of paramount importance to perform the consistency check coherently for those systems.  In particular, for the two-color diquark superfluid and the isospin matter, high quality data from the first-principles lattice simulation are available, whereas the 2SC quark matter calculation could be compared to the future NS observational data.  Our unified treatment must be a useful step toward such efforts.

\section{Formalism}\label{sec:thermo}

\begin{figure}
  \includegraphics[width=0.8\columnwidth]{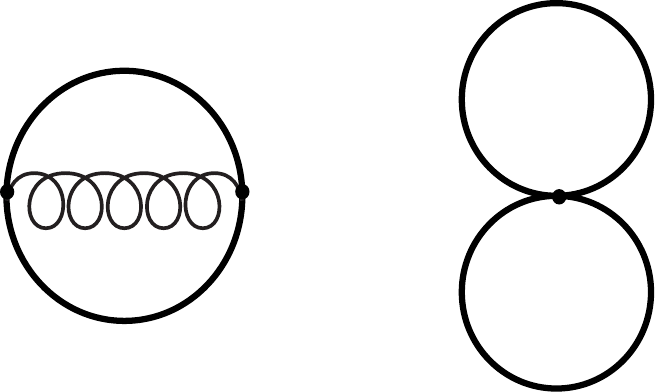}
  \caption{Leading-order diagrams in $\Gamma_2$.  The solid line denotes the quark propagator and the curly line denotes the gluon propagator.  We consider the left diagram in the first part of this work and address the instanton effect from the right diagram in Sec.~\ref{sec:instanton}.}
  \label{fig:diagrams}
\end{figure}

We consider a unified description based on the Cornwall-Jackiw-Tomboulis (CJT) formalism \cite{Cornwall:1974vz, Berges:2004yj} in which the effective action is given by
\begin{align}
  \Omega &= -\frac{1}{2}\Big[\Tr\ln (D^{-1})-\Tr(\Pi D)\Big] \nonumber\\
  &\quad + \eta\Big[\Tr\ln (S^{-1}) - \Tr(\Sigma S)\Big] + \Gamma_2(S, D)\,,
  \label{eq:Omega}
\end{align}
We note that $\eta$ in Eq.~\eqref{eq:Omega} is chosen as $\eta=1/2$ for the calculation of diquark condensate with the Nambu-Gor'kov basis \cite{Nambu:1960tm,gor1958energy} and $\eta=1$ for the calculation of pion condensate.  We use a standard notation;  $S$ and $D$ are the full propagators of quark and gluon, and $\Sigma$ and $\Pi$ are the self-energies of quark and gluon, respectively.  The latter quantities are related to the 2-particle-irreducible diagrams, $\Gamma_2$, through
\begin{align}
    \Sigma &= \frac{1}{\eta}\frac{\delta\Gamma_2}{\delta S}\,, \label{eq:Sigma}\\
    \Pi &= -2\frac{\delta\Gamma_2}{\delta D}\,,\label{eq:Pi}
\end{align}
which corresponds to the stationary condition for the effective action.
The full propagators are written in terms of the free propagators, $S_0$ and $D_0$, and  the self-energies, $\Sigma$ and $\Pi$, through the following Schwinger-Dyson equations:
\begin{align}
    S^{-1}&=S_0^{-1}+\Sigma\,,\label{eq:S}\\
    D^{-1}&=D_0^{-1}+\Pi\,. \label{eq:D}
\end{align}
In this work, we approximate $\Gamma_2$ with truncation up to two types of diagrams as shown in Fig.~\ref{fig:diagrams}.  The left diagram in Fig.~\ref{fig:diagrams} is what is called the sun-set diagram mediated by the one-gluon-exchange interaction, and the right diagram is induced by the instanton interaction as explained below.

Usually, in the pQCD calculation, the non-perturbative contribution from the instanton-induced interaction is not taken into account \cite{Freedman:1976ub, Beluni:1978, Kurkela:2009gj}.  Later, for the demonstration of possible non-perturbative enhancement, we will quantify the instanton effect using the effective interaction for two-flavor matter, i.e.,
\begin{align}
  \mathcal{L}_{\mathrm{inst}} = -\frac{G}{8(\Nc^2-1)}\frac{\Nc^2-1}{\Nc^2} \; 4(\bar{u}\gamma_5 d)(\bar{d}\gamma_5 u)
\end{align}
for the $\bar{q}q$ channel and
\begin{equation}
  \begin{split}
  \mathcal{L}_{\mathrm{inst}} &= -\frac{G}{8(\Nc^2-1)}\frac{\Nc+1}{\Nc}\frac{1}{\xi} \\
  &\quad \times(\bar{\psi}\tau_2\lambda_A\gamma_5 C \bar{\psi}^T)(\psi^T \tau_2\lambda_A C \gamma_5\psi)
  \end{split}
\end{equation}
for the $qq$ channel \cite{Schafer:1996wv, Rapp:1997zu}. Here $\tau_i$'s are Pauli matrices in flavor space, and $\lambda_A$ ($A=2,5,7$) are the anti-symmetric color generators normalized by $\Tr_{\mathrm{color}}(\lambda_A\lambda_B)=\xi \delta_{AB}$, where $\xi=1/2$ in our convention. The charge conjugation matrix is $C=i\gamma^2\gamma^0$.
The strength of the interaction, $G$, is characterized by the instanton density. 
 At the one loop order of the dilute-gas approximation, $G$ or dimensionless coupling, $\bar{G}=G\mu^2$, is given by~\cite{tHooft:1976snw}
\begin{align}
  \bar{G} = C_N \biggl[\frac{8\pi^2}{g(\bar{\Lambda})^2}\biggr]^{2\Nc} \biggl(\frac{\Lambda^2_{\mathrm{QCD}}}{\mu^2}\biggr)^{b_0}\frac{(2\pi)^4}{\Nf^{b_0+1}}\frac{\Gamma(b_0+1)}{2}\,.
  \label{eq:instanton_density}
\end{align}
It should be noted that $\mu$ represents either the quark chemical potential $\muq$ for diquark condensate or the isospin chemical potential $\muiso$ for pion condensate.
Here, $b_0=(11\Nc-2\Nf)/6$ is the constant in the running coupling constant and $C_N$ is a scheme-dependent constant and $\bar{\Lambda}$ is the renormalization scale. Throughout this paper, we adopt the $\overline{\mathrm{MS}}$ scheme, and for details about the corresponding value of $C_N$, see Appendix~\ref{app:inst}.

In order to proceed, we rewrite the full propagators as~\cite{Giannakis:2004xt, Noronha:2006cz} 
\begin{align}
    S(\Delta) &= S_{\mathrm{norm}} + \delta S(\Delta)\,, \\
    D^{-1}(\Delta) &= D^{-1}_{\mathrm{HDL}} + \delta \Pi(\Delta)\,.
\end{align}
The full quark propagator $S$ generally depends on the gap energy $\Delta$.
In this paper, we always refer to $\Delta$ as the gap (energy); $\Delta$ has the dimension of energy, while quark-bilinear condensates have the mass dimension three.
We separate the $\Delta$-independent part as $S_{\mathrm{norm}}=S(\Delta=0)$.  In the above expression, $\delta S$ represents the correction to the quark propagator by $\Delta \neq 0$.
As for the gluon propagator, we consider the separation for not $D(\Delta)$ but $D^{-1}(\Delta)$ for later convenience.  In the above we adapt the Hard-Dense-Loop resummed propagator to approximate $D^{-1} |_{\Delta=0} \simeq D^{-1}_{\mathrm{HDL}}$ \cite{Pisarski:1988vd, Manuel:1995td}.  The resummed propagator is given by
\begin{align}
    D_{\mathrm{HDL}}^{-1}=D_0^{-1}+\Pi_{\mathrm{HDL}}\,.
\end{align}
We give the explicit forms of $\Pi_{\mathrm{HDL}}$ and $D_{\mathrm{HDL}}^{-1}$ in Appendix~\ref{app:notations}.  Plugging these propagators into Eq.~\eqref{eq:Omega}, we obtain the grand potential:
\begin{align}
    \Omega(\Delta) = \Omega_{\mathrm{norm}} + \delta\Omega(\Delta)\,,
\end{align}
where 
\begin{align}
    & \Omega_{\mathrm{norm}} = -\frac{1}{2}\big[\Tr\ln(D_{\mathrm{HDL}}^{-1})-\Tr(\Pi_{\mathrm{HDL}}D_{\mathrm{HDL}})\big] \nonumber\\
    &+\eta \big[\Tr\ln(S_{\mathrm{norm}}^{-1}) \!-\! \Tr(1\!-\!S_0^{-1}S_{\mathrm{norm}})\big]
    + \Gamma_2 |_{\Delta=0} \,.
    \label{eq:omega_n}
\end{align}
We note that the loop corrections are also split as
\begin{equation}
    \Gamma_2 = \Gamma_2|_{\Delta=0} + \delta \Gamma_2\,.
\end{equation}
Technically, $\Gamma_2|_{\Delta=0}$ is diagrammatically calculable in terms of the propagators with $\Delta=0$, i.e., $\Gamma_2(S_{\mathrm{norm}}, D_{\mathrm{HDL}})$.
The remaining part including $\Delta$ dependence reads as:
\begin{align}
    &\delta\Omega(\Delta) = -\frac{1}{2}\big[\Tr\ln(1+\delta\Pi D_{\mathrm{HDL}}) - \Tr(\delta\Pi D_{\mathrm{HDL}})\big] \nonumber\\
    &\: + \eta\bigl[ -\Tr\ln(1+S_{\mathrm{norm}}^{-1}\delta S) + \Tr(S_0^{-1}\delta S) \bigr] + \delta\Gamma_2 \,.
    \label{eq:deltaOmega}
\end{align}

We first address the term, $\Omega_{\mathrm{norm}}$, which gives the pressure in the normal phase.  The weak coupling expansion of Eq.~\eqref{eq:omega_n} with the truncated $\Gamma_2$ given in Fig.~\ref{fig:diagrams} correctly reproduces pQCD up to the NLO results.  It is, therefore, a reasonable approximation to employ the NLO results for the normal phase pressure, that is,
\begin{align}
    p_{\mathrm{norm}} = p_0 \, \gamma_0(g) \,.
    \label{eq:pres_normal}
\end{align}
Here, $p_0$ represents the LO pressure and $\gamma_0(g) = 1-g^2/(2\pi^2)+\mathcal{O}(g^4)$ is the NLO correction~\cite{Freedman:1976ub, Beluni:1978, Kurkela:2009gj}.

Next, let us evaluate $\delta\Omega$ and the pressure induced by $\Delta$.  From now on, we distinguish the state of matter by the subscript $\alpha$ which takes ``2SC'' for the 2SC quark matter, ``2c'' for the two-color diquark superfluid, and ``$\pi$'' for the pion-condensed high-isospin matter.  It should be noted that we use ``pion'' even for high-isospin matter in which the pion no longer exists as a bound state.

In Eq.~\eqref{eq:deltaOmega}, for convenience, we shall refer to the first term as a gluonic part, and to the rest as a fermionic part, and divide the pressure correction to their respective contributions as
\begin{align}
  \delta p_\alpha
  = p_{0} \frac{\Delta_\alpha^2}{\mu^2} \gamma_{1,\alpha}(g)
  = p_{0} \frac{\Delta_\alpha^2}{\mu^2} \bigl[ \gamma_{1,\alpha}^G(g)+\gamma_{1,\alpha}^F(g) \bigr]\,. \label{eq:pres_Gap}
\end{align}
The fermionic part, $\gamma_{1,\alpha}^F$, is derived up to $\mathcal{O}(g)$ in \cite{Fujimoto:2023mvc}:
\begin{align}
  \gamma_{1,\alpha}^F(g)=c^F_{1,\alpha} \bigl[ 1+{c^F_{g,\alpha}}g + \mathcal{O}(g^2) \bigr]\,.
  \label{eq:gammaF}
\end{align}
The values of $c_{1,\alpha}^F$ and $c_{g,\alpha}^F$ for different $\alpha$ are listed in Table.~\ref{tab:coeff}. 
In addition, the $\mathcal{O}(g^2)$ terms from the fermionic-sunset diagram are also evaluated for  2SC matter; see~\cite{, Geissel:2024nmx}. 

\begin{table}
  \begin{tabular}{c|cccc}
   \hline
    $\alpha$~ & ~~$c_{1,\alpha}^F$~ & ~~~$c_{g,\alpha}^F$~~ & ~~~$h_{1,\alpha}$~~~ & $h_{2,\alpha}$ \\
    \hline\hline
    2SC & $3\frac{2}{\Nc}$ & $\frac{1}{3\sqrt{3}}\sqrt{\frac{2\Nc}{\Nc+1}}$ & $\frac{1}{12}\frac{1}{2}\Big(1+\frac{1}{\Nc}\Big)$ & $\frac{2\Nf}{8\Nc(\Nc-1)}$ \\
    2c & $3$ & $\frac{1}{3\sqrt{3}}\sqrt{\frac{2\Nc}{\Nc+1}}$ & $\frac{1}{12}\frac{1}{\Nc}\Big(1+\frac{1}{\Nc}\Big)$ & $\frac{\Nc\Nf}{8\Nc(\Nc-1)}$ \\
    $\pi$ & $3$ & $\frac{1}{3\sqrt{3}}\sqrt{\frac{1}{{C}_{F}}}$ & $\frac{1}{12}{C}_{F}$ & $\frac{2\Nf\Nc}{8\Nc^2}$ \\
    \hline
  \end{tabular}
  \caption{Coefficients in $\gamma_{1,\alpha}^F$ and the gap equation~\eqref{eq:gap_eq} for the 2SC phase ($\alpha=$``2SC''), the diquark superfluid in $\QCtwo$ ($\alpha=$``2c''), and the pion condensation at high isospin density ($\alpha=$``$\pi$'').  For $C_F$, see the expression below Eq.~\eqref{eq:Z}.}
  \label{tab:coeff}
\end{table}

For the color-singlet condensations, namely, $\alpha=$``2c'' and ``$\pi$'', we see that $\gamma_{1,\alpha}^G(g)$ is of $O(g^2)$, so that we can safely neglect such higher-order correction in our approximation.  In contrast, for $\alpha=$``2SC'' with colored diquarks, the magnetic gluons acquire the Meissner mass of the same order as the Debye mass $\propto g\mu$~\cite{Rischke:2000qz, Kojo:2014vja}.  As shown in Appendix~\ref{app:gluon}, this Meissner mass gives a contribution of $O(g\mu^2\Delta^2)$ to the pressure.  For a precise evaluation of this contribution, it is necessary to perform the numerical integration with the momentum-dependent gaps~\cite{Evans:1999at}.  In this paper, for simplicity, we adopt the following expression,
\begin{align}
  \gamma_{1,\mathrm{2SC}}^G (g)=\frac{5\sqrt{2}\pi}{8}g + O(g^2)\,,
  \label{eq:gluon_g_corr}
\end{align}
which is derived by the integration over the restricted region.  This term is of the same order of magnitude as $\gamma_{1,\alpha}^F$ in Eq.~\eqref{eq:gammaF}, and we should keep it.

\section{Behavior of the gaps}\label{sec:gapeq}

Our primary interest lies in the effect of nonzero gaps on the thermodynamic quantities.  Here, we analyze the behavior of $\Delta_{\mathrm{2SC}}$, $\Delta_{\mathrm{2c}}$, and $\Delta_\pi$, respectively.  We can determine $\Delta_\alpha$ by solving the gap equations for respective $\alpha$.  Here, we utilize a generic notation, $\bar{\Delta}_\alpha(\nu)$, where the tilde represents the dimensionless gap normalized by the chemical potential.  Similarly, $\nu$ represents the dimensionless Euclidean energy normalized by the chemical potential.  Dropping the spatial momentum dependence in $\Delta_\alpha$, we arrive at the following gap equation:
\begin{equation}
  \begin{split}
    \bar{\Delta}_{\alpha}(\nu)
    &= \frac{1}{\pi^2}\int_0^{\infty}d\nu' \bigg[ h_{1,\alpha} g^2 \ln \biggl(\frac{b}{\sqrt{|\nu^2-{\nu'}^2|}}\biggr) \\
    &\quad + {h_{2,\alpha}} {\bar{G}}\bigg] Z^2(\nu')\frac{\bar{\Delta}_{\alpha}(\nu')}{\sqrt{{\nu'}^2+\bar{\Delta}_{\alpha}(\nu')^2}}\,,
    \end{split}
    \label{eq:gap_eq}
\end{equation}
where $b=2^8 \pi^4 (2/\Nf)^5 g^{-5}$ \cite{Schafer:1999jg, Pisarski:1999tv}.
We note that the solution of the above gap equation minimizes the effective potential within the approximation made for the derivation of Eq.~\eqref{eq:gap_eq}.  The first logarithmic term corresponds to the sun-set diagram in Fig.~\ref{fig:diagrams} and the second term involving $\bar{G}$ appears from the instanton effect.  We introduce two constants, $h_{1,\alpha}$ and $h_{2,\alpha}$, which depend on the system of our interest; see Table.~\ref{tab:coeff}.
Here, $Z(\nu)$ is the wave-function renormalization factor related to the non-Fermi liquid behavior of quark matter given by~\cite{Manuel:2000mk, Gerhold:2005uu}
\begin{align}
    Z(\nu)^{-1}= 1 + \frac{g^2}{24\pi^2}C_F\ln\biggl(\frac{\pi\bar{m}_{\mathrm{D}}^2}{4\nu^2}\biggr)
    \label{eq:Z}
\end{align}
with $C_F=(\Nc^2-1)/2\Nc$ and the dimensionless Debye mass
$\bar{m}_{\mathrm{D}}^2=(\Nf/2\pi^2) g^2$.

\begin{figure}
  \includegraphics[width=0.95\columnwidth]{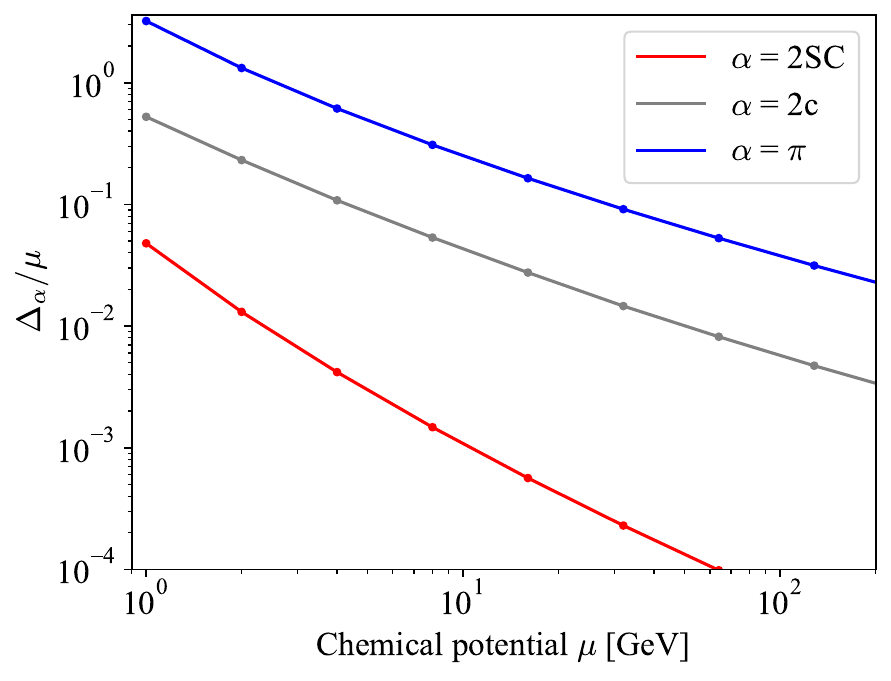}
  \caption{Numerically solved $\Delta_\alpha=\Delta_\alpha(\nu=0)$ from the gap equation~\eqref{eq:gap_eq} where $\bar{\Lambda}=\mu$ is chosen.}
  \label{fig:gap_compare1}
\end{figure}

\begin{figure}
  \includegraphics[width=0.95\columnwidth]{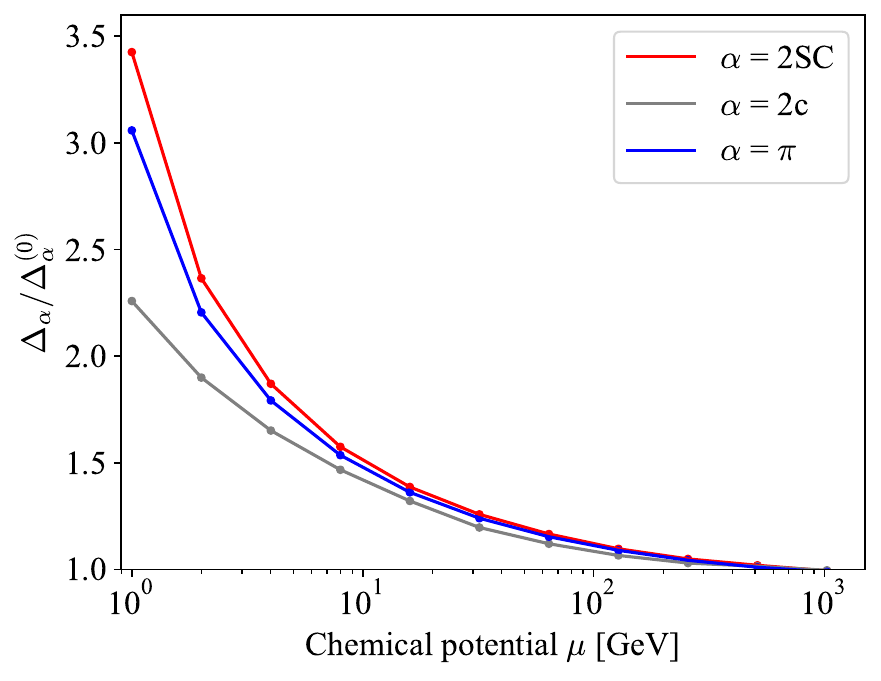}
  \caption{Ratio between the numerically solved $\Delta_\alpha$ and the analytically solved $\Delta_\alpha^{(0)}$.  See Appendix~\ref{app:notations} for the explicit form of $\Delta_\alpha^{(0)}$.}
  \label{fig:gap_compare2}
\end{figure}

We numerically solve the gap equation~\eqref{eq:gap_eq}.  In Figs.~\ref{fig:gap_compare1} and~\ref{fig:gap_compare2}, we show our numerical results without instanton, i.e., the results at $G=0$.  We will later discuss the instanton effect in Sec.~\ref{sec:instanton}.  In solving the gap equation, we introduce a cutoff proportional to the Debye mass, i.e., $Am_D$ with $A$ chosen so that our numerical solution and the analytical solution (see Appendix~\ref{app:notations}) coincide at $\mu=3\;\mathrm{TeV}$.  We note that this cutoff is introduced not to eliminate the ultraviolet (UV) divergence;  the integral in the gap equation~\eqref{eq:gap_eq} has no UV divergence due to the energy dependence in $\Delta_\alpha$.  Nonetheless, the physical cutoff should be imposed to restrict the perturbative formula~\eqref{eq:Z} in the applicable region with $\nu\ll\mu$~\cite{Gerhold:2005uu}.  It should also be noted that the analytical solution assumes an expansion to drop higher-order terms in the gap equation, especially in $Z(\nu)$, while no expansion is employed in our numerical results.

From Fig.~\ref{fig:gap_compare1}, evidently, we find hierarchy, $\Delta_{\pi} > \Delta_{\mathrm{2c}}\gg \Delta_{\mathrm{2SC}}$, and this trend is more prominent at asymptotically high density.  We will come back to this point in later discussions. At high density around $\mu\simeq 1\;\mathrm{TeV}$, the numerical solution asymptotically approaches the analytical solution \cite{Brown:1999aq, Wang:2001aq}, as confirmed in Fig.~\ref{fig:gap_compare2}, though the matching condition is imposed at $\mu=3\;\mathrm{TeV}$ in our prescription. Interestingly, for $\mu\lesssim 10\;\mathrm{GeV}$, it is clear from Fig.~\ref{fig:gap_compare2} that the analytically expressed $\Delta_\alpha^{(0)}$ from the approximate gap equation expanded in $g^2$ underestimate the gaps as compared to the numerically obtained $\Delta_\alpha$ by a factor.  This quantitative enhancement will turn out to be crucial for the corrections in the speed of sound particularly in $\QCtwo$.

\section{Gap effects on the speed of sound}\label{sec:sos}

We present our central results in this section.  Before going to the numerical results, we shall pay attention to the formal consideration for the gap effect on the speed of sound squared, $c_s^2=dp/d\epsilon$ where $\epsilon$ is the internal energy density.  Using Eqs.~\eqref{eq:pres_normal}, and \eqref{eq:pres_Gap}, and the thermodynamic relation, $c_s^2=(\partial p/\partial\mu)/(\mu\partial^2 p/\partial\mu^2)$, we can expand the deviation of $c_s^2$ from the conformal limit as follows:
\begin{align}
  c_s^2 - \frac{1}{3}
  &\simeq -\frac{5}{36}\mu\frac{\partial\gamma_0(g)}{\partial\mu}
    +\frac{\gamma_1(g)}{18} \notag\\
    &\quad \times \biggl\{ 2\frac{\Delta^2}{\mu^2} - \frac{\Delta}{\mu}\frac{\partial\Delta}{\partial\mu} - \Big(\frac{\partial\Delta}{\partial\mu}\Big)^2 - \Delta\frac{\partial^2\Delta}{\partial\mu^2}\biggr\}\,.
  \label{eq:sos_pert_exp}
\end{align}
In this expansion, we kept the leading contributions without and with $\Delta$ using the counting that $\partial\gamma_i(g)/\partial\mu$ is $g^2$ higher order than $\gamma_i(g)$.
The first term is the normal phase NLO correction without $\Delta$ which is always negative.  We should recall that $\mu$-dependence in $\gamma_0$ appears from $-g(\mu)^2$ which gets larger with increasing $\mu$, resulting in $-\partial\gamma_0(g)/\partial\mu > 0$.  This is why the pQCD usually predicts the speed of sound approaching the conformal limit from below~\cite{Kojo:2021wax, Geissel:2024nmx}.
The rest is the correction from the gaps.  If we drop the last term $\propto \partial^2\Delta/\partial\mu^2$, we see that three terms are factored as $(\Delta/\mu - \partial\Delta/\partial\mu)(2\Delta/\mu + \partial\Delta/\partial\mu)$.  From this form, it is obvious that the speed of sound could rise above the conformal limit when $|{\Delta}/{\mu}| \gg |{\partial\Delta}/{\partial\mu}|$.  In other words, even if the condensate substantially develops at some density, the whole correction to the speed of sound could even be negative for $|\partial\Delta/\partial\mu| \gg |\Delta/\mu|$~\cite{Geissel:2024nmx}.  It is important to note that both signs of $\partial\Delta/\partial\mu$ can make the whole combination negative.  Later, we will discuss the trace anomaly and point out that the structure is very different there.

\begin{figure}
  \includegraphics[width=0.95\columnwidth]{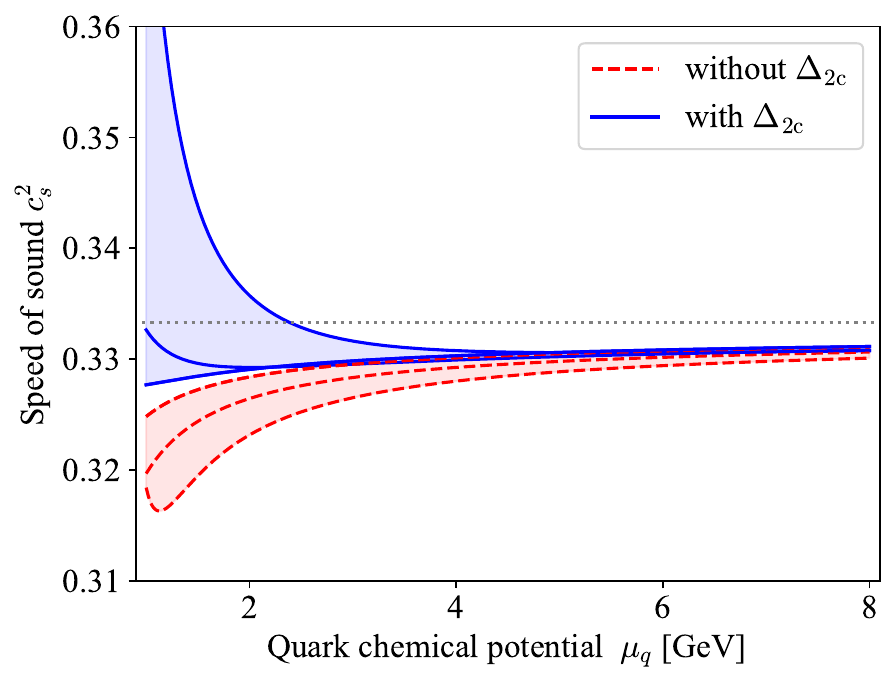}
\caption{Speed of sound for the diquark superfluid in $\QCtwo$ without instanton effect.  The red dashed lines and blue solid lines represent the results without and with the gap effect, respectively, where three variants are for $X=1, 2, 4$ with $\overline{\Lambda}=X\mu$.}
\label{fig:sos2}
\end{figure}

\begin{figure}
  \includegraphics[width=0.95\columnwidth]{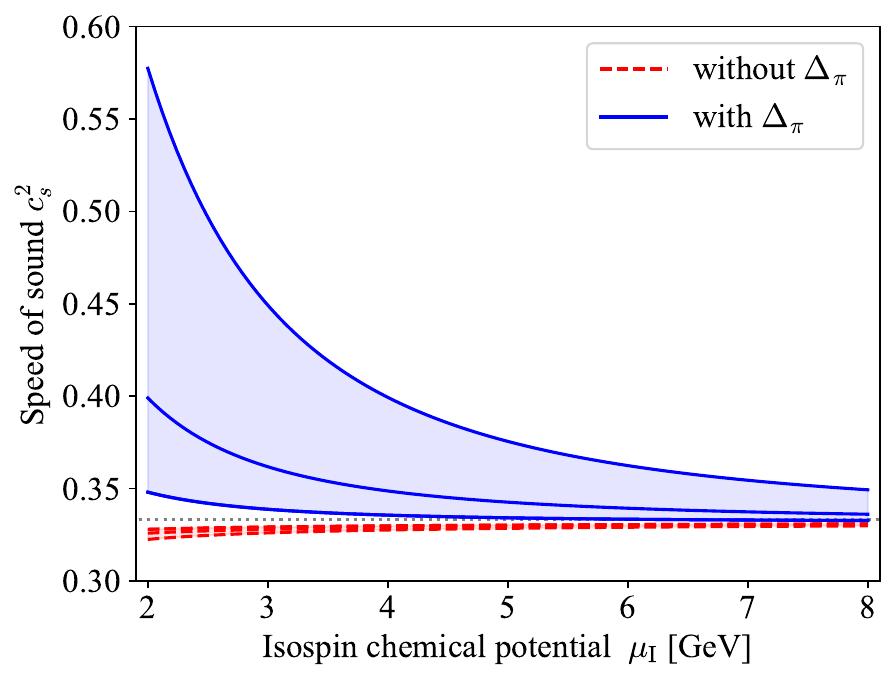}
\caption{Speed of sound for the pion-condensed high-isospin matter without instanton effect.  The red dashed lines and blue solid lines represent the results without and with the gap effect, respectively, where three variants are for $X=1, 2, 4$ with $\overline{\Lambda}=X\mu$.}
\label{fig:sos3}
\end{figure}

\begin{figure}
  \includegraphics[width=0.95\columnwidth]{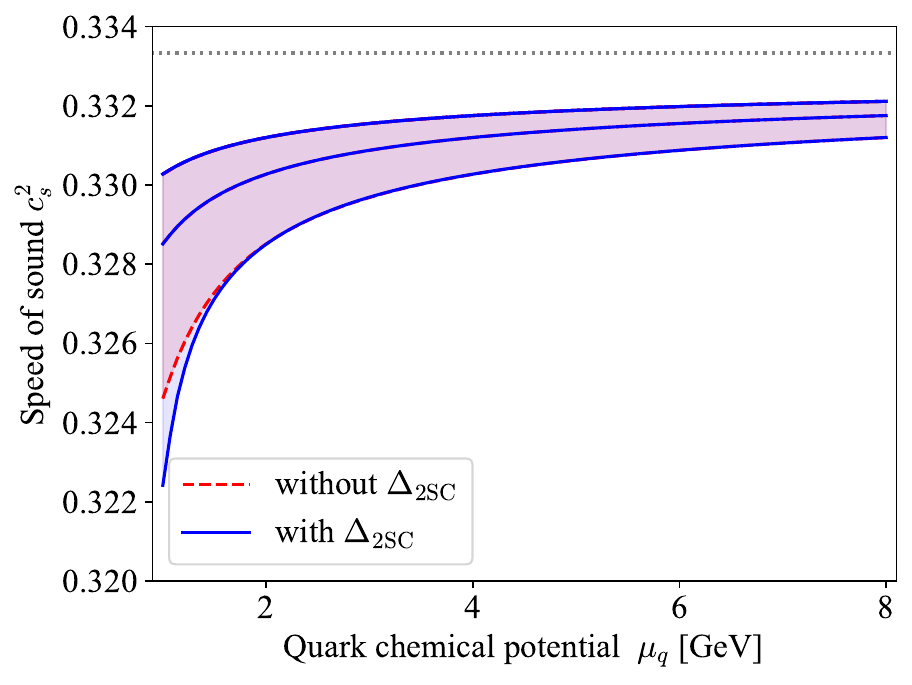}
\caption{Speed of sound for the 2SC quark matter without instanton effect.  The red dashed lines and blue solid lines represent the results without and with the gap effect, respectively, where three variants are for $X=1, 2, 4$ with $\overline{\Lambda}=X\mu$.}
\label{fig:sos1}
\end{figure}

In Figs.~\ref{fig:sos2},~\ref{fig:sos3}, and \ref{fig:sos1}, we plot the speed of sound for the diquark superfluid in $\QCtwo$, the pion-condensed high-isospin matter, and the 2SC quark matter, respectively, without instanton effect.
Apparently, the gap effect in Fig.~\ref{fig:sos1} is exceptionally minor, thus we shall discuss the results in Figs.~\ref{fig:sos2} and~\ref{fig:sos3} first.

In both cases of the diquark superfluid in  $\QCtwo$ and the pion-condensed high-isospin matter, the speed of sound without $\Delta$ remains smaller than the conformal limit, while the gap corrections make $c_s^2$ exceed the conformal limit in the intermediate density region~\cite{Fujimoto:2023mvc}.  As observed in Figs.~\ref{fig:sos2} and \ref{fig:sos3}, $\Delta_\pi$ causes a drastic rise in the speed of sound as compared to $\Delta_{\mathrm{2c}}$.  This difference is simply attributed to the size of $\Delta_\alpha$.  In fact, the exponents are different as $\Delta_{\mathrm{2c}}\sim \mu\exp[-2\pi^2/g(\mu)]$ and $\Delta_\pi\sim \mu_{I}\exp[-3\pi^2/2g(\mu_I)]$.  This difference in the exponential factor results in larger $\Delta_\pi$ by an order of magnitude than $\Delta_{\mathrm{2c}}$ for the same isospin/quark chemical potential.   The size difference of $\Delta_\alpha$ makes a sharp contrast in the high-density region where the na\"{i}ve perturbation is expected to be valid.  In the high-isospin case, $\Delta_\pi$ is so large that the speed of sound can be considerably modified even at $\muiso \gtrsim 2\;\text{GeV}$.  Recently, several lattice QCD simulations have carefully estimated the behavior of $c_s^2$ at high isospin density~\cite{Brandt:2022hwy, Abbott:2023coj}.  Their results also suggest $c_s^2$ above the conformal value for isospin matter up to $\muiso\sim 2\;\text{GeV}$, which may be consistently connected to our results.  In the $\QCtwo$ case, on the other hand, it is still difficult to make a direct comparison between our results and lattice $\QCtwo$ simulation~\cite{Iida:2022hyy, Iida:2024irv}, which is partly because the fermion masses are larger than physical values and partly because the simulation does not reach sufficient precision at high density where pQCD should work.  Nevertheless, the lattice $\QCtwo$ data are consistent with our results within large uncertainty.  Unlike the isospin case in Fig.~\ref{fig:sos3}, our perturbation calculations show good convergence around $\muq\gtrsim 3\;\text{GeV}$, and thus our results in Fig.~\ref{fig:sos2} should be reliable enough to provide the baseline for the validity check of the lattice $\QCtwo$ simulations.

We next turn to the results for the 2SC quark matter in Fig.~\ref{fig:sos1}.  Apparently, the gap correction to $c_s^2$ is a minor effect; it is positive but invisibly small at high density.  In the low density region, the gap correction becomes negative, and $c_s^2$  deviates farther below the conformal limit.  This small and even opposite correction is explained from Eq.~\eqref{eq:sos_pert_exp} with $|\partial\Delta/\partial\mu| \sim |\Delta/\mu|$.  This suppression due to $|\partial\Delta/\partial\mu| \sim |\Delta/\mu|$ may or may not change with higher-order corrections to the gap equation.  We will revisit this point in the next section.

We make a comment about related non-perturbative studies.  According to the recent analysis of the quark-meson-diquark model and fRG~\cite{Braun:2022jme, Geissel:2024nmx, Andersen:2024qus}, it is a favorable scenario that the speed of sound exceeds the conformal limit in the 2SC quark matter at some intermediate densities.  In our results, in contrast, the speed of sound does not reach the conformal limit, which seems to conflict with phenomenology.  However, recalling that our perturbative results for the diquark superfluid in $\QCtwo$ and the pion-condensed high-isospin matter are consistent with lattice QCD simulations, it is conceivable enough that that our 2SC results in Fig.~\ref{fig:sos1} can be reliable as long as the uncertainty band is narrow.  Therefore, to reconcile our results and others, one likely scenario is that the speed of sound in the 2SC quark matter stays below the conformal limit for $\muq\gtrsim 2\;\text{GeV}$ and it rapidly increases in the non-perturbative regime at lower densities~\cite{Kurkela:2024xfh}.  We will discuss a possible mechanism for this rapid increase of the speed of sound in Sec.~\ref{sec:implication}.

\section{Gap effects on the trace anomaly}\label{trace}

We shall discuss the behavior of the trace anomaly.  We first define the normalized trace anomaly as
$\mathfrak{T} = 1/3 - p/\epsilon$ and then we expand $\mathfrak{T}$ in the same way as Eq.~\eqref{eq:sos_pert_exp} to find:
\begin{align}
  \mathfrak{T}
  & = \frac{1}{3} - \frac{p}{\epsilon} \nonumber\\
  & \simeq \frac{\mu}{9}\,\frac{\partial\gamma_0(g)}{\partial\mu} - \frac{2}{9}\gamma_1(g) \frac{\Delta}{\mu}\Big(\frac{\Delta}{\mu}-\frac{\partial\Delta}{\partial\mu}\Big) \,.
  \label{eq:trace_anomaly}
\end{align}
It is important to note that in this case, again, $\Delta$-induced corrections appear in the form of the difference of the $\Delta$ and its derivative. For $\Delta\sim \mathcal{O}(\Lambda_{\mathrm{QCD}})$ and $\Delta/\mu\gg\partial\Delta/\partial\mu$, only the first term $\propto (\Delta/\mu)^2$ becomes dominant and the $\Delta$ correction should be negative.

More interestingly, in Eq.~\eqref{eq:trace_anomaly}, the derivative contribution appears with the non-derivative one in the form of $\Delta/\mu - \partial\Delta/\partial\mu$, so that the relative sign is crucial.  When the gaps grow up rapidly as the density get lowered, we have $\partial\Delta/\partial\mu < 0$ and there is no cancellation between $\Delta/\mu$ and $\partial\Delta/\partial\mu$.  This makes a sharp contrast to the behavior of the speed of sound;  $(\partial\Delta/\partial\mu)^2$ cancels $(\Delta/\mu)^2$ regardless of the sign of the derivative.

\begin{figure}
\includegraphics[width=0.95\columnwidth]{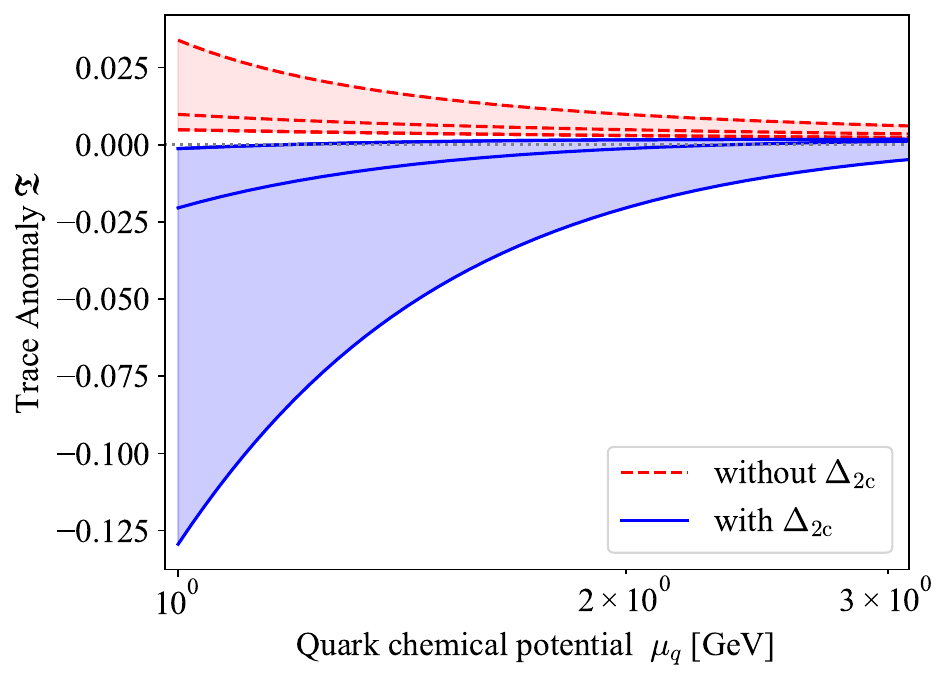}
\caption{Comparison of the normalized trace anomaly without $\Delta_{\mathrm{2c}}$ (red dashed lines) and with $\Delta_{\mathrm{2c}}$ (blue solid lines).}
\label{fig:2c_ta}
\end{figure}

\begin{figure}
\includegraphics[width=0.95\columnwidth]{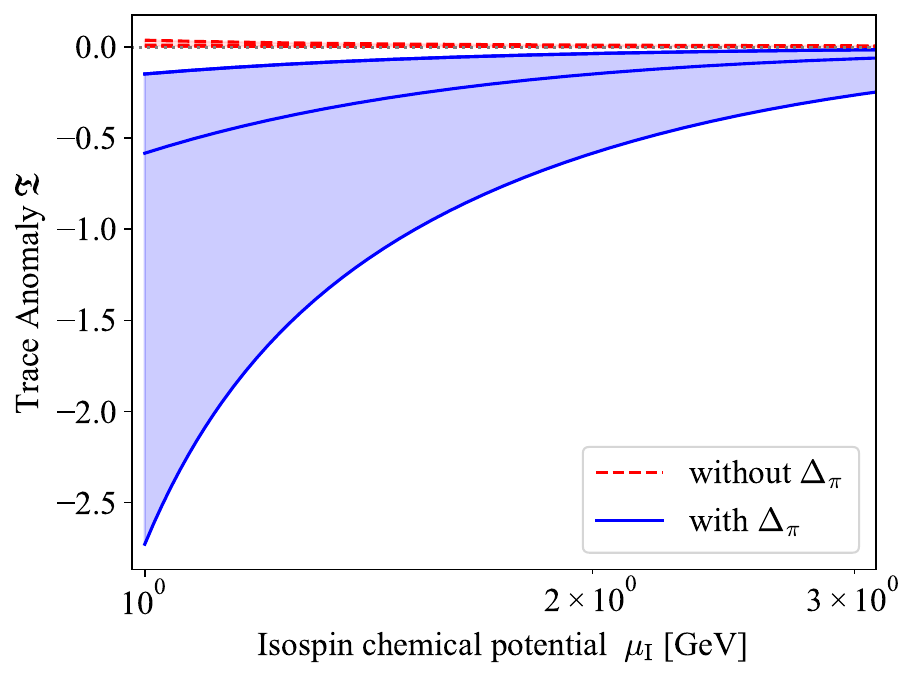}
\caption{Comparison of the normalized trace anomaly without $\Delta_{\pi}$ (red dashed lines) and with $\Delta_{\pi}$ (blue solid lines).}
\label{fig:pi_ta}
\end{figure}

\begin{figure}
\includegraphics[width=0.95\columnwidth]{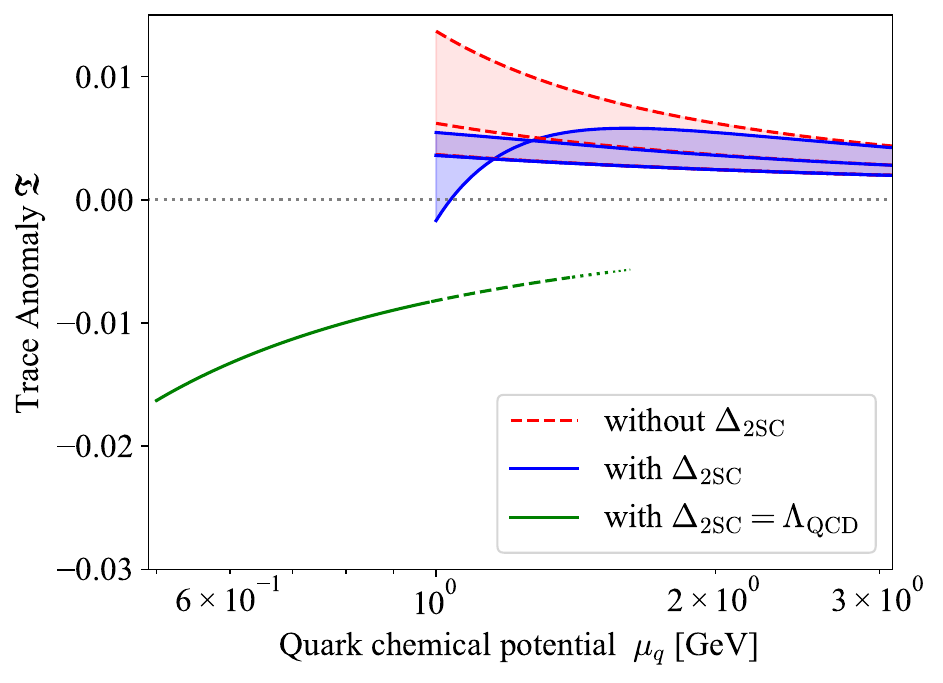}
\caption{Comparison of the normalized trace anomaly without $\Delta_{\mathrm{2SC}}$ (red dashed lines) and with $\Delta_{\mathrm{2SC}}$ (blue solid lines).
The green line represents the results with $\Delta_{\mathrm{2SC}}=\Lambda_{\mathrm{QCD}}$.}
\label{fig:2sc_ta}
\end{figure}

In Figs.~\ref{fig:2c_ta}, \ref{fig:pi_ta}, and \ref{fig:2sc_ta}, we show our perturbative results for the normalized trace anomaly $\mathfrak{T}$ for the diquark superfluid in $\QCtwo$, the pion-condensed high-isospin matter, and the 2SC quark matter, respectively.  Clearly, the trace anomaly is significantly modified by the effects of $\Delta_\alpha\neq 0$ and it tends to go negative, as suggested from the NS data analysis~\cite{Fujimoto:2022ohj} and confirmed in lattice QCD calculations~\cite{Iida:2022hyy,Brandt:2022hwy,Abbott:2023coj,Iida:2024irv}.
The results in Figs.~\ref{fig:2c_ta} and \ref{fig:pi_ta} are not very surprising in view of the corresponding enhancement in the speed of sound as presented in Figs.~\ref{fig:sos2} and \ref{fig:sos3}.

It is highly nontrivial that $\Delta_{\mathrm{2SC}}$ substantially affects the normalized trace anomaly around $\muq\simeq 1\;\text{GeV}$ as seen from Fig.~\ref{fig:2sc_ta}, while the change in the speed of sound in Fig.~\ref{fig:sos3} is only minor.  Although the uncertainty band is large, $\mathfrak{T}$ could become negative by the effects of $\Delta_{\mathrm{2SC}}$ within the perturbative framework.  At the same time, the crossing behavior of the uncertainty band by $X=1, 2, 4$ indicates that the validity of the perturbative analysis may break down at lower densities.  We cannot extrapolate our perturbative results to the non-perturbative regions at lower densities, and nevertheless, it would be instructive to see the impact of the non-perturbative effects qualitatively.  To this end, instead of using the gap equation solution, we set $\Delta_{\mathrm{2SC}}=\Lambda_{\mathrm{QCD}}$ by hand and calculate $\mathfrak{T}$ using the expressions in terms of $\Delta_{\mathrm{2SC}}$.  Then, the normalized trace anomaly is further pushed down negatively, as shown by the green line in Fig.~\ref{fig:2sc_ta}.  This is a desirable output, and one may think that some non-perturbative mechanism to enhance $\Delta_{\mathrm{2SC}}$ must be implemented in the calculations, but it is a subtle question how to keep consistency as we discuss in the next section.

\section{Instanton-induced enhancement}\label{sec:instanton}

So far, we have seen the results without the instanton-induced interaction.  As long as we consider the sunset diagram in Fig.~\ref{fig:diagrams}, we find that the speed of sound in the 2SC quark matter is hardly modified by the diquark condensate, while the trace anomaly can receive non-negligible contributions.  The instanton excitation is generally Debye-screened by finite density~\cite{Pisarski:1980md, Rapp:1997zu}, so that our perturbative treatment should be reasonable if the density is high enough.

In the non-perturbative regime at intermediate densities around $\muq\simeq \Lambda_{\mathrm{QCD}}$, however, the screening effect may not be dominant, and the instanton-induced interaction could play an important role in phenomenology~\cite{Schafer:1996wv, Rapp:1997zu, Alford:1998mk, Rajagopal:2000wf, Fukushima:2010bq}.
Actually, $G$ in Eq.~\eqref{eq:instanton_density}, the strength of the instanton-induced interaction, has parametric dependence on $\mu$ as $\mu^{-2}(\Lambda_{\mathrm{QCD}}^2/\mu^2)^{b_0}\ln(\mu/\bar{\Lambda})^{2\Nc}$, which is not exponential but power-law suppression.  Thus, it is a legitimate strategy toward the non-perturbative regime to take account of the instanton-induced interaction as a first step.

\begin{figure}
  \includegraphics[width=0.95\columnwidth]{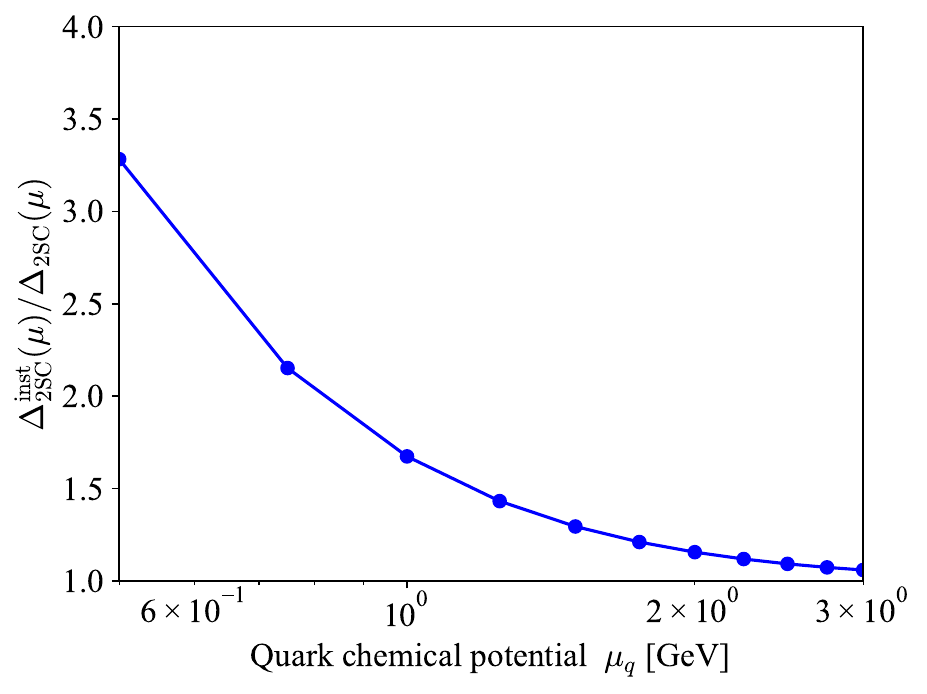}
  \caption{Diquark gaps$\Delta_{\mathrm{2SC}}^{\mathrm{inst}}$ with the instanton-induced interactions and $\Delta_{\mathrm{2SC}}$ as functions of the quark chemical potential.  The renormalization scale, $\bar{\Lambda}=\muq$, is chosen.}
  \label{fig:2sc_gap}
\end{figure}

\begin{figure}
  \includegraphics[width=0.95\columnwidth]{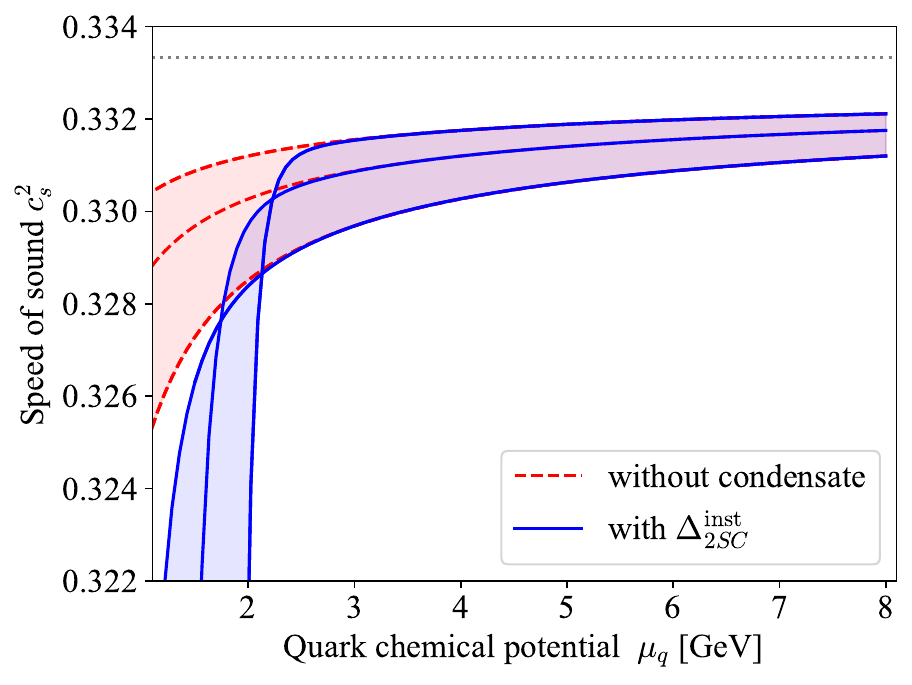}
  \caption{Speed of sound calculated with and without the effects of $\Delta_{\mathrm{2SC}}^{\mathrm{inst}}$.}
  \label{fig:2sc_inst}
\end{figure}

In Fig.~\ref{fig:2sc_gap}, as in previous studies~\cite{Schafer:2004yx}, we numerically confirm that the instanton effects cause enhancement in $\Delta_{\mathrm{2SC}}$ by a factor $\sim 2$-$3$ for $\muq\lesssim 1\;\text{GeV}$.  We do not show a figure but the trace anomaly is negatively shifted as expected.
However, enhanced $\Delta_{\mathrm{2SC}}$ does not necessarily lead to desirable behavior of the speed of sound.  In fact, as shown in Fig.~\ref{fig:2sc_inst}, the speed of sound is shifted downward from the conformal limit for $\muq\lesssim 2\;\text{GeV}$.
We can understand these results from the expression of the speed of sound in terms of $\Delta$ and $\partial\Delta/\partial\mu$ in Eq.~\eqref{eq:sos_pert_exp}.  Since $\Delta_{\mathrm{2SC}}^{\mathrm{inst}}$ exhibits steeper increase, the instanton effects may well lead to $|\partial\Delta/\partial\mu| > |\Delta/\mu|$, which flips the sign of the correction to suppress the speed of sound.
We have checked that this decreasing behavior of the speed of sound is common also for the diquark superfluid in $\QCtwo$ and the pion-condensed high-isospin matter.  In these cases, the suppression of the speed of sound is disfavored by the lattice QCD results~\cite{Iida:2022hyy,Brandt:2022hwy,Abbott:2023coj,Iida:2024irv}.  Therefore, we conclude that the (dilute) instanton effects should not be the leading non-perturbative correction; either the instanton-induced interactions are still Debye-screened and/or other non-perturbative effects overcome.

\section{A scenario for non-perturbative behavior of the gaps}\label{sec:implication}

\begin{figure}
  \includegraphics[width=0.85\columnwidth]{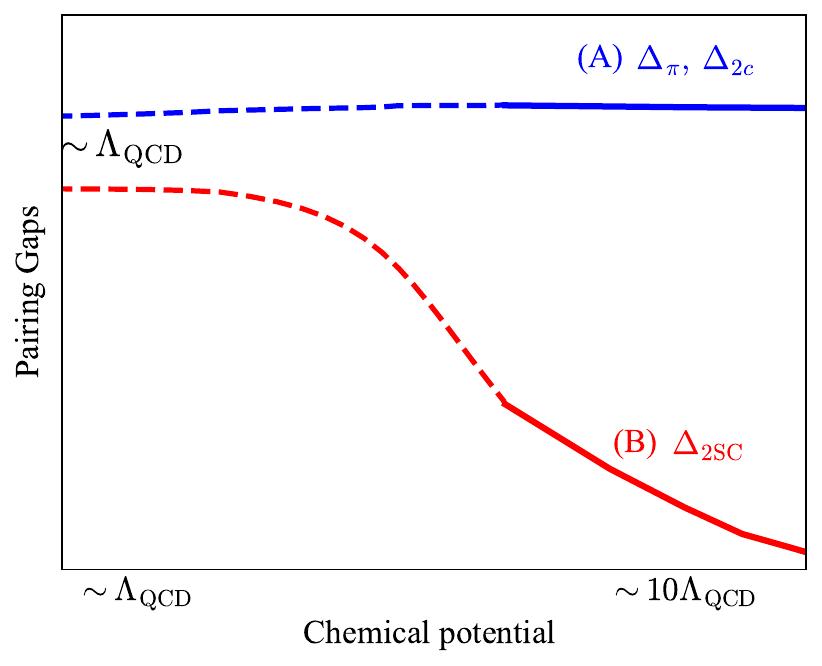}
  \caption{Schematic illustration for our scenario of the gap behavior in the non-perturbative regime with logarithmic scales.
    [A: diquark superfluid in $\QCtwo$ and pion-condensed high-isospin matter] The gap energy of $\mathcal{O}(\Lambda_{\mathrm{QCD}})$ at high density with $\mu/\Lambda_{\mathrm{QCD}} \sim \mathcal{O}(10)$ (blue solid line) remains almost unchanged at lower density with $\mu/\Lambda_{\mathrm{QCD}}\sim \mathcal{O}(1)$ (blue dashed line).
    [B: 2SC quark matter] The gap much smaller than $\Lambda_{\mathrm{QCD}}$ at high density (red solid line) rapidly increases at lower density and then saturated at some value of $\mathcal{O}(\Lambda_{\mathrm{QCD}})$ (red dashed line).}
  \label{fig:gap_concept}
\end{figure}

In Sec.~\ref{sec:sos}, we have seen that the gaps significantly enhance the speed of sound in the cases of the diquark superfluid in $\QCtwo$ and pion-condensed high-isospin matter.  Interestingly, we have confirmed that the speed of sound can exceed the conformal limit within the density range where pQCD is reliable, whereas the gap effect is negligibly small for the 2SC quark matter.  As discussed before, it is the density dependence of the gap that makes this difference.  We shall take a closer look at the cancellation mechanism and describe a likely scenario for non-perturbative behavior.

Figure.~\ref{fig:gap_concept} schematically summarizes our speculation on the non-perturbative gap behavior.  For $\QCtwo$ and high-isospin matter, the gap takes a value of $\Delta\sim \Lambda_{\mathrm{QCD}}$ even at very high density with $\mu/\Lambda_{\mathrm{QCD}} \sim \mathcal{O}(10)$.  We have found that the $\mu$-dependence is mild, so that $|\Delta/\mu|\gg|\partial\Delta/\partial\mu|$.  This behavior can be extrapolated toward lower density regions where the chiral perturbation theory makes model-independent predictions.  There, the gap is still $\Delta\sim f_\pi\sim\Lambda_{\mathrm{QCD}}$ at the intermediate density with $\mu/\Lambda_{\mathrm{QCD}}\sim\mathcal{O}(1)$.  As a first approximation, therefore, the gap magnitude can be regarded as constant as shown by the dashed blue line in Fig.~\ref{fig:gap_concept} and the derivative terms can be dropped in Eq.~\eqref{eq:sos_pert_exp}.  Under this simplification, only the $(\Delta/\mu)^2$ term of $\mathcal{O}(1)$ for $\mu\sim\Lambda_{\mathrm{QCD}}$ contributes to raise the speed of sound.
On the other hand, in the case of the 2SC quark matter, the diquark gap turns out to be much smaller than $\Lambda_{\mathrm{QCD}}$ at high density with $\mu/\Lambda_{\mathrm{QCD}}\sim \mathcal{O}(10)$.  Because the gap rapidly increases at lower densities, $|\partial\Delta/\partial\mu| \sim |\Delta/\mu|$ is realized.  The derivative term $(\partial\Delta/\partial\mu)^2$ overcomes the positive contribution of $(\Delta/\mu)^2$ in Eq.~\eqref{eq:sos_pert_exp}, and then the speed of sound decreases.  The microscopic origin of these differences lies in the strength of the one-gluon-exchange (OGE) interaction.  Since the color-singlet channel is the most attractive in the gluon interactions, the order of the OGE strength is (isospin) $>$ ($\QCtwo$) $\gg$ (2SC)~\cite{Son:1998uk, Son:2000by, Kanazawa:2009ks}.  Because of the strong suppression of the 2SC gap at high density, the gap should rapidly be rising at lower densities.

It is conceivable that in the intermediate density with $\mu\sim\Lambda_{\mathrm{QCD}}$, the diquark gap can increase to $\mathcal{O}(\Lambda_{\mathrm{QCD}})$ due to the non-perturbative interactions.  In this regime, all the non-perturbative mass scales are characterized by $\Lambda_{\mathrm{QCD}}$, and it is likely that the diquark gap is saturated and its increasing rate becomes small, i.e., $\Delta/\mu\gg\partial\Delta/\partial\mu$. Once this is realized, $(\Delta/\mu)^2$ in Eq.~\eqref{eq:sos_pert_exp} would raise the speed of sound.  Such behavior of the diquark gap should be manifested in the quark-meson-diquark model. Recent calculations~\cite{Andersen:2024qus} predict that the speed of sound begins to increase at $\mu\sim 1.5\;\text{GeV}$, and the maximum value is $c_s^2\sim0.38$ at $\mu\sim 0.4\;\text{GeV}$. This maximum value is quantitatively smaller than the observational data from the neutron stars and the lattice calculations, but the qualitative behavior is reasonable.

So far, we have focused on the 2SC phase only.  In the case of three-flavor quark matter, the strength of diquark attraction is the same for color superconductivity in the three-flavor case.  In fact, the perturbative gap energy obtained in the CFL phase of three-flavor quark matter takes the same exponent and the difference is found only in the overall factors~\cite{Alford:2007xm}.
Therefore, the CFL gap, $\Delta_{\mathrm{CFL}}$, should be of the same order as the 2SC gap and the impact of $\Delta_{\mathrm{CFL}}$ to the speed of sound and the trace anomaly is expected to be similar to what we have seen for the 2SC quark matter in the present study.
Interestingly, the NS data analysis also suggests that $\Delta_{\mathrm{CFL}}$ is likely to be as small as the perturbative estimate at high density~\cite{Kurkela:2024xfh, Fujimoto:2024pcd}.  Therefore, if the peak in the speed of sound and the negative trace anomaly are attributed to the color-superconducting gap, some non-perturbative mechanism as speculated above should be necessary.  Probably, $\Delta_{\mathrm{CFL}}$ is taken over by $\Delta_{\mathrm{2SC}}$ in the intermediate density region.

\section{Summary}\label{sec:summary}

We examined the role of the condensates in thermodynamics in two-flavor dense quark matter under the various extreme conditions such as high baryon and isospin density.
We employed a unified treatment of the gap equation based on the perturbative calculation.  Specifically, we extensively studied the peak structure in the speed of sound squared, $c_s^2$, and the behavior of the normalized trace anomaly, $\mathfrak{T}$, at densities enough to justify the perturbative estimates. To this end, we revisited some approximations in the gap equation and the pressure correction by the gaps.
We confirmed that the numerically solved condensates make $c_s^2$ exceed the conformal limit and $\mathfrak{T}$ go down to the negative values in the dense $\QCtwo$ and high-isospin cases. Comparing these results with the lattice simulation results, we found reasonable agreement, which supports the reliability of our computation. 

Armed with these consistent results, we proceeded to the same analysis for the two-flavor color-superconducting, 2SC, matter.
We then found that the gap correction to $c_s^2$ is minor in this case. The difference between the 2SC case and others is attributed to the magnitude of the gap; i.e., $\Delta_{\pi} > \Delta_{\mathrm{2c}}\gg \Delta_{\mathrm{2SC}}$ follows from numerical constants in the gap equation. More interestingly, we demonstrated that $\mathfrak{T}$ receives sizable gap corrections and approaches negative. This change in $\mathfrak{T}$ is along the consistent direction with the observation, while $c_s^2$ decreases and does not agree with what is expected from the observation. Such a quantitative difference between $\mathfrak{T}$ and $c_s^2$ arises from the interplay between the gap $\Delta/\mu$ and its derivative $\partial{\Delta}/\partial\mu$. According to Eqs.~\eqref{eq:sos_pert_exp} and \eqref{eq:trace_anomaly}, as long as $|\partial{\Delta}/\partial\mu|\ll |\Delta/\mu|$, a large enough $\Delta/\mu$ increases $c_s^2$ and decreases $\mathfrak{T}$ as na\"{i}vely expected.  However, in the case of the 2SC quark matter, $\Delta_{\mathrm{2SC}}$ rapidly increases as $\mu$ becomes smaller. Therefore, the $(\partial{\Delta}/\partial\mu)^2$ term overwhelms the $(\Delta/\mu)^2$ term for $c_s^2$ in Eq.~\eqref{eq:sos_pert_exp}. In contrast, such cancellation does not occur for $\mathfrak{T}$ in Eq.~\eqref{eq:trace_anomaly}.

Here, in the present study, we focused on the two-flavor matter.  As the key equations~\eqref{eq:sos_pert_exp} and \eqref{eq:trace_anomaly} do not depend on $\Nf$, our conclusions about qualitative findings are expected to be applicable to the three-flavor matter.  Thus, only quantitative differences appear from some numerical constants in the gap equation.

In addition, we added the instanton-induced interaction to the gap equation in order to examine a possible non-perturbative correction at intermediate density. We adopted the strength of the interaction based on the approximation of the dilute instanton gas.  Contrary to the recent lattice simulation results, we found that the instanton-induced interaction rapidly reduces $c_s^2$.   This unwanted behavior was seen also for the two-color diquark superfluid and the pion-condensed high-isospin matter.
Therefore, our results indicate that the instanton effects might have been overestimated~\cite{Pisarski:2024esv}.  Alternatively, some other non-perturbative effects could override the results or further suppress the instanton density in the intermediate density region.
These topics deserve more investigation in the future.

\begin{acknowledgments}
The authors thank
Etsuko~Itou,
Lorenz~von~Smekal, Toru~Kojo, and
Yuya~Tanizaki
for useful discussions.
This work was supported by Japan Society for the Promotion of Science
(JSPS) KAKENHI Grant Nos.\
22H01216 and 22H05118 (KF), 24KJ0985 (SM).
\end{acknowledgments}

\appendix
\section{Useful formulae}
\label{app:notations}
We use the one loop coupling constant
\begin{align}
    g^2(\overline{\Lambda})=\frac{(4\pi)^2}{\beta_0 \ln({\overline{\Lambda}^2}/{\Lambda^2_{\mathrm{QCD}})}},
\end{align}
where $\beta_0 = (11N_{\mathrm{c}}-2N_{\mathrm{f}})/3$. We choose $\Lambda_{\mathrm{QCD}}=0.2$ GeV. 
The renormalization-scale ambiguity originates from the coupling constant $g(\overline{\Lambda})$. 
In this paper, we adopt the one loop coupling constant. It was also confirmed that there was no significant difference in the conclusion of the main text when the higher-loop coupling constant is used. 
The renormalization scale is chosen as $\overline{\Lambda}=X\mu$ with $X=1,2,4$. 

For the momentum scale in the gapped phase, the Hard-Dense-Loop propagator must be used. For $k_0\sim \Delta$ and $k\sim m_D$, these propagators are given by~\cite{Manuel:1995td}
\begin{align}
    D_{{E}, \mathrm{HDL}}(k_0,k)&\simeq\frac{1}{k^2+m_D^2},\\
    D_{{M}, \mathrm{HDL}}(k_0,k)&\simeq\frac{1}{k^2+\frac{\pi}{4}m_D^2\frac{|k_0|}{k}} .
\end{align}
The subscription E and M means ``electric'' and ``magnetic'' respectively.

Also, the analytical results of  $\Delta^{(0)}_{\alpha}$ are summarized as~\cite{Brown:1999aq, Wang:2001aq, Fujimoto:2023mvc, Kanazawa:2009ks}:
\begin{align}
    \Delta^{(0)}_{\mathrm{2SC}}&=\mu_q \ 512\pi^4 g^{-5}e^{-\frac{4+\pi^2}{8}}\exp\Big(-\frac{{3\pi^2}}{\sqrt{2}g}\Big),\\
    \Delta^{(0)}_{\mathrm{2c}}&=\mu_q \ 512\pi^4 g^{-5}e^{-\frac{4+\pi^2}{16}}\exp\Big(-\frac{{2\pi^2}}{g}\Big),\\
    \Delta^{(0)}_{\mathrm{\pi}}&=\mu_{\mathrm{I}} \ 1024\pi^4 g^{-5}e^{-\frac{4+\pi^2}{16}}\exp\Big(-\frac{{3\pi^2}}{2g}\Big).
\end{align}

\section{Instanton density in \texorpdfstring{$\overline{\rm MS}$}{TEXT} scheme}
\label{app:inst}
The instanton density depends on the renormalization scheme only through the numerical factor $C_N$ which is expressed as
\begin{equation}
    C_N = \frac{C_1\, e^{-C_2 \Nc}\, e^{C_3\Nf}}{(\Nc-1)!(\Nc-2)!}\,.
\end{equation}
This factor including $C_1$, $C_2$, $C_3$ was calculated by 't~Hooft for the first time by the Pauli-Villars (PV) regularization~\cite{tHooft:1976snw}:
\begin{align}
C_1^{\mathrm{PV}}&=2e^{5/6}/\pi^2\,,\\
C_2^{\mathrm{PV}}&=\frac{5}{3}\ln 2-\frac{17}{36}+\frac{1}{3}(\ln 2\pi+\gamma_{E})-\frac{2}{\pi^2}\zeta'(2)\,,\\
C_3^{\mathrm{PV}}&=-\frac{1}{3}\ln 2-\frac{17}{36}+\frac{1}{3}(\ln 2\pi+\gamma_{E})-\frac{2}{\pi^2}\zeta'(2),
\end{align} 
where $\gamma_{E}\simeq 0.5772$ is the Euler-Mascheroni constant. 

For the translation to the dimensional regularization, we need the translation formula of $\Lambda_{\mathrm{QCD}}$ among different renormalization schemes. The formula from PV to MS was also derived in Ref.~\cite{tHooft:1976snw} with minor errors.  Numerical factors after the errors are corrected can be found in Ref.~\cite{Hasenfratz:1981tw}.  The translated coefficients used in this paper are
\begin{align}
C_1^{\overline{\mathrm{MS}}}&=C_1^{\mathrm{PV}}\simeq 0.466\,,\\
C_2^{\overline{\mathrm{MS}}}&=C_2^{\mathrm{PV}}-\frac{1}{6}\simeq 1.511\,,\\
C_3^{\overline{\mathrm{MS}}}&=C_3^{\mathrm{PV}}-\frac{1}{33}\simeq 0.261.
\end{align} 

\section{Gluon contribution to the pressure}
\label{app:gluon}
We give a detailed derivation of Eq.~\eqref{eq:gluon_g_corr}. Our starting point is the diquark correction to the one-loop gluonic CJT potential:
\begin{align}
\frac{1}{2}\sum_{a=4}^{8}\int\frac{d^4k}{(2\pi)^4}\Big[\ln(1+\delta\Pi^a_M D^a_{M,\mathrm{HDL}})-\delta\Pi^a_M D^a_{M,\mathrm{HDL}}\Big],
\label{eq:gluon_free_ene}
\end{align}
where $\delta\Pi\equiv \Pi-\Pi(\Delta=0)$, and the subscription M means the magnetic gluon. The above expression involves the summation over only $a=4$-$8$ with the magnetic components.  This is because corresponding $\delta\Pi$'s for the electric gluons and for the magnetic gluons of unbroken parts ($a=1$-$3$) are negligibly small~\cite{Kojo:2014vja}.
Also, for technical simplicity, we assume all gluons have the same self-energy as the 8th-gluon: $\delta\Pi^a_M=\delta\Pi^8_M$. In fact, this approximation overestimates the Meissner mass~\cite{Rischke:2000qz}.

Under these approximations, we can analytically calculate Eq.~\eqref{eq:gluon_free_ene}. Since the dominant region of the Hard-Dense-Loop propagator is $k_0\sim\Delta,\ k \sim m_{D}$, we can approximate Eq.~\eqref{eq:gluon_free_ene} as follows:
\begin{align}
\frac{5\Delta}{2}\int\frac{d^3k}{(2\pi)^4}\Big[\ln\Big(1+\frac{\delta\Pi^8_{M,0}}{k^2} \Big)-\frac{\delta\Pi^8_{M,0}}{k^2}\Big],
\label{eq: gluon free ene}
\end{align}
where $\delta\Pi^a_{M,0}\equiv\delta\Pi^a_{M}(k_0=0)$.  We can easily evaluate the self-energy as
\begin{align}
&\delta\Pi^a_{M,0}(k)\nonumber\\&= \frac{N_Fg^2\mu^2}{24\pi^2}\int dx_0\int_{-1}^{1}dx \frac{(3-x)}{\sqrt{x_0^2+1}(x_0^2+1+\frac{k^2x^2}{4})},
\label{eq: gluon free ene2}
\end{align}
where all the momenta are normalized by $\Delta$. From this expression, we obtain the following asymptotics: $\delta\Pi^a_{M,0}(k)=\frac{N_Fg^2\mu^2}{2\pi^2}$ for $k\lesssim 1$, and $\delta\Pi^a_{M,0}(k)=\frac{N_Fg^2\mu^2}{4k}$ for $k\gtrsim 1$. 

Separating the integral~\eqref{eq: gluon free ene} into the two regions where these approximations are valid and performing the remaining integrals rigorously, we recover Eq.~\eqref{eq:gluon_g_corr} as
\begin{align}
&\frac{5\Delta}{2(4\pi^3)}\Big(\int_0^1+\int_1^\infty\Big)dk\ k^2\Big[\ln(1+\frac{\delta\Pi}{k^2} )-\frac{\delta\Pi}{k^2}\Big]\nonumber\\
&\simeq \frac{5}{48\pi^3}g^2\mu^2\Delta^2\ln(m_D^2/\Delta^2)\simeq-\frac{5\sqrt{2}}{16\pi}g\mu^2\Delta^2.
\end{align}

\bibliography{diquark}
\bibliographystyle{apsrev4-2}

\end{document}